\documentstyle[12pt]{article}

\def\hybrid{\topmargin -20pt	\oddsidemargin 0pt
	\headheight 0pt	\headsep 0pt
        \textwidth 6.35in
        \textheight 9.65in
	\marginparwidth .875in
	\parskip 5pt plus 1pt	\jot = 1.5ex}

 \catcode`\@=11

\@addtoreset{equation}{subsection}
\@addtoreset{footnote}{section}
\def\theequation{\thesubsection.\arabic{equation}}

\newtoks\@stequation

\def\subequations{\refstepcounter{equation}%
  \edef\@savedequation{\the\c@equation}%
  \@stequation=\expandafter{\theequation}
  \edef\@savedtheequation{\the\@stequation}
  \edef\oldtheequation{\theequation}%
  \setcounter{equation}{0}%
  \def\theequation{\oldtheequation\alph{equation}}}

\def\endsubequations{\setcounter{equation}{\@savedequation}%
  \@stequation=\expandafter{\@savedtheequation}%
  \edef\theequation{\the\@stequation}\global\@ignoretrue
  \vspace*{-12pt} \\}

\def\pr{Phys. Rev. \/}
\def\np{Nucl. Phys. \/}
\def\pl{Phys. Lett. \/}

\hybrid
\def\baselinestretch{1.2}
\catcode`\@=11
\def\marginnote#1{}
\catcode`\@=11
\def\section{\@startsection {section}{1}{0pt}{-3.5ex plus -1ex minus
 -.2ex}{2.3ex plus .2ex}{\raggedright\large\bf}}
\catcode`\@=12
\newskip\humongous \humongous=0pt plus 1000pt minus 1000pt

\newif\ifdtup

\def\be{\begin{equation}}
\def\ee{\end{equation}}
\def\ba{\begin{eqnarray}}
\def\ea{\end{eqnarray}}
\def\bs{\begin{subequations}}
\def\es{\end{subequations}}

\def\p{\partial}

\def\t{\tau}
\def\tbb{\bar\tau}
\def\R{{\cal{R}}}
\def\Q{{\cal{Q}}}
\def\RR{{\rm I\!R}}

\begin{document}
\renewcommand{\theequation}{\thesection.\arabic{equation}}
\newcommand{\beq}{\begin{equation}}
\newcommand{\eeq}[1]{\label{#1}\end{equation}}
\newcommand{\ber}{\begin{eqnarray}}
\newcommand{\eer}[1]{\label{#1}\end{eqnarray}}
\begin{titlepage}
\begin{center}

\hfill CERN-TH.7472/94\\
\hfill LPTENS-94/36\\
\hfill hep-th/9501020\\

\vskip .2in

{\large \bf Infrared Regularization of Superstring Theory and the
One-Loop Calculation of Coupling Constants}
\vskip .4in

{\bf Elias Kiritsis and Costas Kounnas\footnote{On leave from Ecole
Normale Sup\'erieure, 24 rue Lhomond, F-75231, Paris, Cedex 05,
FRANCE.}}\\
\vskip
 .3in

{\em Theory Division, CERN,\\ CH-1211,
Geneva 23, SWITZERLAND} \footnote{e-mail addresses:
KIRITSIS,KOUNNAS@NXTH04.CERN.CH}\\

\vskip .3in

\end{center}

\vskip .2in

\begin{center} {\bf ABSTRACT } \end{center}
\begin{quotation}\noindent

Infrared regularized versions of 4-D N=1 superstring ground states
are constructed by curving the spacetime. A similar regularization
can
be performed in field theory. For the IR regularized string  ground
states
we derive the exact one-loop effective action for non-zero U(1) or
chromo-magnetic fields as well as gravitational and axionic-dilatonic
fields.
This effective action is IR and UV finite.
Thus,  the one-loop corrections to
all couplings (gravitational, gauge and Yukawas) are
unambiguously computed.
These corrections are necessary for quantitative string
superunification
predictions at low energies.
The one-loop corrections to the couplings are also found to satisfy
Infrared Flow Equations.

\end{quotation}
\vskip 1.0cm
\vskip 2.cm
CERN-TH.7472/94 \\
December 1994\\
\end{titlepage}
\vfill
\eject
\def\baselinestretch{1.2}
\baselineskip 16 pt
\noindent
\section{Introduction}
\setcounter{equation}{0}

The four-dimensional superstring solutions in a flat background
\cite{cand}-\cite{gepner}
 define, at  low energy, effective supergravity theories
\cite{effcl}-\cite{ant}.
 A class of them successfully extends
the validity of the standard model up to the string scale,
$M_{str}$.
 The first main property of superstrings is that they are
ultraviolet-finite theories (at least perturbatively). Their
second important
property is that  they unify gravity with all other interactions.
This unification does not include  only the gauge interactions, but
also the  Yukawa  ones as well as the interactions among the scalars.
This String Hyper Unification (SHU) happens  at  large energy scales
$E_t\sim {\cal O}(M_{str})\sim 10^{17}~$GeV. At this energy scale,
however,
the first excited string states become important and thus the whole
effective low energy field theory picture breaks
down \cite{n4kounnas}-\cite{kktopol}. Indeed, the
effective field theory of strings is valid only for  $E_t \ll
M_{str}$ by means of
the ${\cal O}(E_t/M_{str})^2$ expansion. It is then necessary to
evolve the SHU predictions to a lower scale $M_U < M_{str}$ where
the  effective field theory picture makes sense. Then, at $M_U$, any
string solution provides  non-trivial relations between  the gauge
and
Yukawa couplings, which can be written as\footnote{The logarithmic
part
was calculated for the first time in string theory in \cite{min}.}

\begin{equation}
\frac{k_i}{\alpha_i(M_U)}=\frac{k_j}{\alpha_j(M_U)}+\Delta_{ij}(M_U).
\label{shu}
\end{equation}

The above relation looks very similar to the well-known unification
condition
in Supersymmetric Grand Unified Theories (SuSy-GUTs) where the
unification scale is about $M_U\sim 10^{16}~$GeV and
$\Delta_{ij}(M_U)=0$ in the ${\bar {DR}}$ renormalization scheme; in
SuSy-GUTs the normalization constants $k_i$ are fixed $only$ for the
gauge couplings ($k_1=k_2=k_3=1$, $k_{em}=\frac{3}{8}$), but there
are
no relations among gauge and Yukawa couplings at all. In string
effective theories, however, the normalization constants ($k_i$) are
known for both gauge and Yukawa interactions. Furthermore,
$\Delta_{ij}(M_U)$ are calculable $finite$ quantities for any
particular string solution. Thus, the predictability of a given
string solution is extended for all low energy coupling constants
${\alpha_i(M_Z)}$ once the string-induced corrections
$\Delta_{ij}(M_U)$ are determined.

 This determination  however, requests string computations which
 we did not know, up to now, how to perform in generality. It turns
out
that
$\Delta_{ij}(M_U)$ are non-trivial functions of the vacuum
expectation values  of  some gauge singlet fields
\cite{moduli,dfkz,ant},
$~~\langle T_A\rangle =t_A$, the so-called moduli (the moduli fields
are flat
directions at the string classical level and they remain flat in
string perturbation theory, in the exact supersymmetric limit).
The $\Delta_{ij}(t_A)$ are target space duality invariant functions,
which depend on the particular
string solution.  Partial results for
$\Delta_{ij}$ exist \cite{moduli,dfkz,ant} in the exact
supersymmetric limit in many string
solutions based on orbifold
\cite{orbifold} and fermionic constructions \cite{abk4d}.
As we will see later $\Delta_{ij}$
are, in principle,   well defined
calculable quantities once we perform our calculations  at the string
level where all interactions including gravity are  consistently
defined. The full string corrections   to the coupling constant
unification, $\Delta_{ij}(M_U)$, as well as the string corrections
associated to the soft supersymmetry-breaking parameters

\centerline{$m_0$, $m_{1/2}$, $A$, $B$ and $\mu$, at $M_U$,}

\noindent
are of main importance, since they fix  the strength of the gauge and
Yukawa interactions, the full spectrum of the
supersymmetric particles as well as the Higgs and the top-quark
masses at the low energy range $M_Z\leq E_{t}\leq {\cal O}(1)$ TeV.

In the case where supersymmetry is broken \cite{gcsbr,ssbr}
only
semi-quantitative results can be obtained at present; a much more
detailed
study and understanding are  necessary which is related to the
structure of soft breaking terms after the assumed supersymmetry
breaking \cite{fkpz}.

The main obstruction in determining the exact form of the string
radiative corrections $\Delta_{ij}(M_U)$ is strongly related to the
infrared divergences of the $\langle [F^a_{\mu\nu}]^2\rangle $
two-point correlation function in superstring theory. In field
theory, we
can avoid this problem using off-shell calculations. In first
quantized string theory we cannot do that since we do not know how to
go off-shell.
Even in field theory there are problems in defining an infrared
regulator for chiral fermions especially in the presence of
spacetime supersymmetry.

In \cite{cw} it was suggested to use a specific spacetime with
negative curvature in order to achieve consistent regularization in
the infrared. The proposed  curved space however is not useful for
string applications since it does not correspond to an exact
super-string solution.

Recently, exact and stable superstring solutions
have been constructed  using special four-dimensional spaces as
superconformal building blocks with
${\hat c}=4$ and $N=4$ superconformal
symmetry \cite{n4kounnas,worm}. The full spectrum of string
excitations for the superstring solutions based on those
four-dimensional subspaces, can be derived using the techniques
developed in \cite{worm}. The main characteristic property of these
solutions
is the existence of a mass gap, which is proportional
to the curvature of the non-trivial  four-dimensional spacetime.
Comparing the spectrum in  a flat background with that in curved
space we observe a shifting of all massless states by an amount
proportional to the spacetime curvature, $\Delta m^2=Q^2/4=\mu^2/2$,
where $Q$ is the Liouville background charge and $\mu$ is the IR
cutoff. What is
also interesting is that the shifted spectrum in curved space is
equal for bosons and fermions due to the existence of a new
space-time supersymmetry defined in curved spacetime
\cite{n4kounnas,worm}. Therefore, our curved spacetime
infrared
regularization is consistent with supersymmetry and can be
used either in field theory or string theory.

In section 2 we define the four-dimensional superconformal system
responsible for the IR cutoff and give
the modular-invariant partition function for some symmetric orbifold
ground states of the string.
In section 3 we show how we can deform the theory  consistently, by
switching  on a non-zero gauge field strength background $\langle
F^a_{\mu\nu}F_a^{\mu\nu}\rangle =F^2$ or a gravitational one,
$\langle R_{\mu\nu\rho\sigma}R^{\mu\nu\rho\sigma}\rangle=\R^2$ and
obtain the $exact$
regularized partition function $Z(\mu,F,\R)$.
Our method of constructing this effective action automatically
takes into
account the back-reaction of the other background fields; stated
otherwise, the perturbation that turns on the constant gauge field
strength or curvature background is an exact (1,1) integrable
perturbation. The
second derivative with respect to $F$ of our deformed partition
function
$\p^2 Z(\mu,F,\R)/\p F^2$ for $F,\R=0$ defines without any infrared
ambiguities the
complete  string one-loop corrections to the gauge coupling
constants.
In the $\mu\to 0$ limit we recover the known partial results
\cite{moduli,dfkz,ant}.
A preliminary version of our results has appeared in \cite{trieste}.

\section{Regulating the Infrared}
\setcounter{equation}{0}

Any 4-D string solution that can be used to describe particle
physics is composed from a 4-D flat spacetime CFT (with $c=(6,4)$)
which provides the universal degrees of freedom (graviton,
antisymmetric
tensor and dilaton)
and some internal CFT (with $c=(9,22)$) which provides the various
particle degrees of freedom (gauge fields, fermions, scalars).

We would like to  regularize the IR by turning on background fields
associated to the universal degrees of freedom ($G_{\mu\nu}$,
$B_{\mu\nu}$, $\Phi$) so that it can be used for 4-D string ground
states with arbitrary particle content.
This will be done by replacing the 4-D flat spacetime CFT with
another CFT
which however has to satisfy the following constraints:

{\bf 1.} The string spectrum must have a mass gap $\mu^2$.
          In particular, chiral fermions should be regulated
consistently.

{\bf 2.} We should be able to take the limit $\mu^2\to 0$.

{\bf 3.} It should have $c=(6,4)$ so that it can be coupled to any
         internal CFT with $c=(9,22)$.

{\bf 4.} It should preserve as many spacetime supersymmetries of the
original theory, as possible.

{\bf 5.} We should be able to calculate the regulated quantities
relevant for
the effective field theory.

{\bf 6.} Vertices for spacetime fields (like $F_{\mu\nu}^{a}$) should
be
          well defined operators on the world-sheet.

{\bf 7.} The theory should be modular invariant (which guarantees the
absence
of anomalies).

{\bf 8.} Such a regularization should be possible also at the
effective field theory level. In this way, calculations in the
fundamental theory can be matched
without any ambiguity to those of the effective field theory.

Requirements {\bf 3} and {\bf 4} imply that the 4-D CFT should have
$N=4$ superconformal
symmetry\footnote{It is possible to have higher superconformal
symmetry
but we know of no example that regulates the IR.}.
If we need to regulate an N=1 spacetime supersymmetric ground state
the N=4 requirement can be dropped and N=2 is sufficient.
In this case one can use a 4-D CFT with $c=(6+\epsilon,4+\epsilon)$
and an internal
CFT with $c=(6-\epsilon,22-\epsilon)$. This would come close to the
dimensional regularization of IR divergences used in field theory.
However we have good indications that in the limit $\epsilon\to 0$
the internal theory
decompactifies so we will not consider this possibility further.

There are many N=4 CFTs \cite{kikoulu} that can regulate the IR but
if
we insist on requirement {\bf 5}, then we obtain the following list
of candidates
\cite{kprn4,n4kounnas}:

\ba
{\bf I.}& W_{k}^{(4)} &\equiv U(1)_{Q}\otimes SU(2)_{k_1}
\nonumber\\
{\bf II.}& C^{(4)}_k &\equiv [SU(2)/U(1)]_{k}\otimes
U(1)_{R}\otimes U(1)_{Q}\nonumber\\
{\bf III.}& \Delta^{(4)}_k(A) &\equiv [SU(2)/U(1)]_k
\otimes [SL(2,R)/U(1)_A]_{k+4}\nonumber\\
{\bf IV.}& \Delta^{(4)}_k(V) &\equiv [SU(2)/U(1)]_k
\otimes [SL(2,R)/U(1)_V]_{k+4}\nonumber
\ea

\noindent
and their N=4 preserving continuous deformations.
The CFTs above (and their supersymmetric deformations) are
constructed
out of conformal subsystems whose characters are known
\cite{kac,nonc}
The N=4 superconformal symmetry plays an important role since it
indicates the appropriate modular invariant combinations of
characters for these systems \cite{n4kounnas,worm}.

In this work we will use system {\bf I} but similar considerations
can be advanced for the other systems.

The background charge $Q$ in cases {\bf I} and {\bf II} is related to
the level $k$
due to the $N=4$ algebra, $Q=\sqrt{2/(k+2)}$ and guarantees that
${\hat c}=4$ for any value of $k$.

In the limit of weak curvature (large $k$) the  $W_{k}^{(4)}$ space
can  be interpreted as a topologically non-trivial four-dimensional
manifold of the form $\RR \otimes S^3$. The underlying superconformal
field theory associated
to $W_{k}^{(4)}$ includes a supersymmetric $SU(2)_{k}$ WZW model
describing the three coordinates of $S^3$ as well as a non-compact
dimension with a background charge, describing the scale factor of
the sphere \cite{n4kounnas,worm}. Furthermore this space
admits two covariantly constant
spinors and, therefore, respects up to two space-time supersymmetries
(in the heterotic case)
consistently with the $N=4$ world-sheet symmetry
\cite{wormclas,n4kounnas,worm}.
The explicit representation of the desired $N=4$ algebra is derived
in \cite{kprn4} and \cite{n4kounnas}, while the target space
interpretation
as a four-dimensional semi-wormhole space is given in
\cite{wormclas}.

The basic rules of construction in curved spacetime are similar to
that of the orbifold construction \cite{orbifold}, the free 2-d
fermionic constructions \cite{abk4d},  and the Gepner construction
\cite{gepner} where one combines in a modular-invariant way the
world-sheet degrees of freedom in a way consistent with unitarity and
spin-statistics of  the string spectrum.

Our regulated string ground state is of the form
$W^{(4)}_k\otimes K^{(6)}$, where $K^{(6)}$
is any appropriate internal CFT.
To be explicit, we choose this CFT to be
one of the symmetric orbifold models, used in $(2,2)$
compactifications, although as it will become obvious, this can be
done
for any ``solvable" internal CFT.

Since the world-sheet fermions of the  $W^{(4)}_k$ superconformal
system are free and since the $K^{(6)}$ internal theory is the same
as
in the $\RR^4\otimes K^{(6)}$  4-D superstring solutions, we can
easily
obtain the
partition function of $W^{(4)}_k\otimes K^{(6)}$, $Z^{W}$, for $k$
even, in
terms of that of
$\RR^4\otimes K^{(6)}$, $Z^{F}$:
\begin{equation}
Z^{W}[\mu,\tau, \bar{\tau}]=[\Gamma(SU(2)_k)(\tau, {\bar
\tau})]~Z^F[\tau, {\bar \tau}],
\label{zwf1}
\end{equation}
where $\Gamma (SU(2)_k)$ is nothing but the contribution to the
partition function of the  bosonic coordinates $X^{\mu}$ of the
curved background $W^{(4)}$ divided by the contribution of the four
free coordinates of the 4-D flat space,
\begin{equation}
\Gamma (SU(2)_k)={1\over 2}[({\rm Im}\tau)^{\frac{1}{2}}
\eta(\tau){\bar
\eta}({\bar \tau})]^{3}~~\sum_{a,b=0}^{1}Z^{SU(2)}[^{a}_{b}]
{}.
\label{zwf2}
\end{equation}
\be
Z^{SU(2)}[^a_b]=e^{-i\pi k ab/2}\sum_{l=0}^ke^{i\pi
bl}\chi_l(\tau){\bar \chi}_{l+a(k-2l)}({\bar
\tau})
\ee
where $\chi_l(\tau)$ are the characters of  $SU(2)_k$ (see for
example \cite{su2ch}) and the integer $l$ is equal to twice  the
$SU(2)$ spin  $l=2j$.
It is necessary to use this orbifoldized version of $SU(2)_{k}$
in order to project out negative norm states of the $N=4$
superconformal
representations \cite{worm}.
Note that this factorized form is valid for any 4-D ground state
which has  $N\leq 2$ spacetime supersymmetry in flat space.
This is due to the fact that the $W^{(4)}$ space has two covariantly
constant spinors.
If the original flat space background has $N=4$ spacetime
supersymmetry
then this is broken to $N=2$ via the coupling of SU(2) spin with the
internal
manifold. In such a case eq. ({\ref{zwf1}) changes and its new form
will be presented elsewhere.
Also note that even though the vaccum amplitude ({\ref{zwf1}) has a
factorized form (before the $\tau$ integration) this does not imply
that the one-loop corrections to couplings have a similar factorized
form.
In particular as we will see later on, the one-loop correction to the
$R^2$ coupling as well as the corrections to gauge couplings for
non-supersymmetric
ground states are not factorized.

To obtain the above formula we have used the continuous series of
unitary representations
of the Liouville characters \cite{worm} which are generated by the
lowest-weight  operators,
\begin{equation}
e^{\beta X_L}\ ;\quad \beta=-\frac{1}{2}Q +ip\ ,
\label{contchar}
\end{equation}
having positive conformal weights $h_p=Q^2/8+p^2/2$.
The fixed imaginary part in the momentum $iQ/2$ of the plane waves
is due to the non-trivial dilaton motion.

As a particular example we give below the partition function of the
$Z_2 \otimes Z_2$ symmetric orbifolds \cite{orbifold,abk4d},
$W^{(4)}_k \otimes T^{(6)}$/$(Z^{2}\otimes Z^{2})$, for type-II and
heterotic constructions:
$$
Z^W_{II}[\mu;\tau,{\bar \tau}]~=~ {\Gamma(SU(2)_{k})\over {\rm Im}\t
{}~\eta^2\bar\eta^2}\times
{1\over 16}\sum_{\alpha,\beta,{\bar \alpha},{\bar
\beta}=0}^{1}\sum_{h_1,g_1,h_2,g_2}
Z_1[^{h_1}_{g_1}]Z_2[^{h_2}_{g_2}]Z_3[^{-h_1-h_2}_{-g_1-g_2}]\times
$$
\be
(-)^{\alpha+\beta+\alpha\beta}
\frac{\vartheta[^{\alpha}_{\beta}]}{\eta}
\frac{\vartheta[^{\alpha+h_1}_{\beta+g_1}]}{\eta}
\frac{\vartheta[^{\alpha+h_2}_{\beta+g_2}]}{\eta}
\frac{\vartheta[^{\alpha-h_1-h_2}_{\beta-g_1-g_2}]}{\eta}
{}\times (-)^{{\bar\alpha}+{\bar\beta}+\bar\alpha\bar\beta}
\frac{{\bar\vartheta}[^{{\bar\alpha}}_{{\bar\beta}}]}
{{\bar\eta}}
\frac{{\bar\vartheta}[^{{\bar\alpha}+h_1}_{{\bar\beta}+g_1}]}
{{\bar\eta}}
\frac{{\bar\vartheta}[^{{\bar\alpha}+h_2}_{{\bar\beta}+g_2}]}
{{\bar\eta}}
\frac{{\bar\vartheta}[^{{\bar\alpha}-h_1-h_2}_{{\bar\beta}-g_1-g_2}]}
{{\bar\eta}}\label{type2}
\end{equation}
where $Z_i[^{h_i}_{g_i}]$ in (\ref{type2}) stands for the partition
function of two twisted bosons with twists ($h_i,g_i)$. The untwisted
part $Z_i[^{0}_{0}]$
is equal to the moduli-dependent two-dimensional lattice
$\Gamma(2,2)[T_i,U_i]$/$(\eta{\bar\eta})^2$.
The definition of the $\vartheta$-function we use is
\be
\vartheta[^{a}_{b}](v|\tau)=\sum_{n\in
Z}e^{i\pi\tau(n+a/2)^2+2i\pi(n+a/2)(v+b/2)}
\ee
In the heterotic case, a modular-invariant partition function  can be
easily obtained using the heterotic map \cite{llsmap,gepner}. It
consists in
replacing in (\ref{type2}) the $O(2)$ characters associated to the
right-moving fermionic coordinates ${\bar\Psi}^{\mu}$, with the
characters of either $O(10)\otimes E_8$:
\begin{equation}
(-)^{{\bar\alpha}+{\bar\beta}+\bar\alpha\bar\beta}\;
\frac{{\bar\vartheta}[^{\bar\alpha}_{\bar\beta}]}{{\bar\eta}}
\rightarrow
\frac{{\bar\vartheta}[^{\bar\alpha}_{\bar\beta}]^{5}}{{\bar\eta}^5}
\;
{1\over
2}\sum_{\gamma,\delta}\frac{{\bar\vartheta
[^{\gamma}_{\delta}]}^{8}}{{\bar\eta}^8}\label{pfheta}
\end{equation}
or $O(26)$:
\begin{equation}
(-)^{{\bar\alpha}+{\bar\beta}+\bar\alpha\bar\beta}\;
\frac{{\bar\vartheta}[^{\bar\alpha}_{\bar\beta}]}{{\bar\eta}}
\rightarrow
\frac{{\bar\vartheta[^{\bar\alpha}_{\bar\beta}]}^{13}}
{{\bar\eta}^{13}}.
\label{pfhetb}
\end{equation}
Using the map above, the heterotic partition function  with
$E_{8}\otimes E_{6}$ unbroken gauge group is:

$$
Z^W_{het}[\mu; \tau,\tbb]={\Gamma(SU(2)_{k})\over {\rm Im}\t
{}~\eta^2\bar\eta^2}\times
{1\over 16}
\sum_{\alpha,\beta,{\bar\alpha},{\bar\beta}=0}^{1}
\sum_{h_1,g_1,h_2,g_2}
{Z_1[^{h_1}_{g_1}]Z_2[^{h_2}_{g_2}]Z_3[^{-h_1-h_2}_{-g_1-g_2}]}\times
$$

\be
(-)^{\alpha+\beta+\alpha\beta}
\frac{\vartheta[^{\alpha}_{\beta}]}{\eta}
\frac{\vartheta[^{\alpha+h_1}_{\beta+g_1}]}{\eta}
\frac{\vartheta[^{\alpha+h_2}_{\beta+g_2}]}{\eta}
\frac{\vartheta[^{\alpha-h_1-h_2}_{\beta-g_1-g_2}]}{\eta}\times
{1\over
2}\sum_{\gamma,\delta}\frac{{\bar\vartheta}
[^{\gamma}_{\delta}]^8}{{\bar\eta}^8}\frac{{\bar\vartheta}
[^{\bar\alpha}_{\bar\beta}]^5}{{\bar\eta}^5}
\frac{{\bar\vartheta}[^{{\bar\alpha}+h_1}_{{\bar\beta}+g_1}]}
{{\bar\eta}}
\frac{{\bar\vartheta}[^{{\bar\alpha}+h_2}_{{\bar\beta}+g_2}]}
{{\bar\eta}}
\frac{{\bar\vartheta}
[^{{\bar\alpha}-h_1-h_2}_{{\bar\beta}-g_1-g_2}]}
{{\bar\eta}}
\label{het}
\end{equation}

The mass spectrum of bosons and fermions in both the  heterotic and
type-II constructions is degenerate due to the existence of
space-time
supersymmetry defined in the $W^{(4)}_k$ background. The heterotic
constructions are  $N=1$ spacetime supersymmetric while in the
type-II construction one obtains  $N=2$ supersymmetric solutions.

The boson (or fermion) spectrum is obtained  by setting to $+1$ (or
to $-1$)
the statistical factor,
$(-)^{\alpha+\beta+{\bar\alpha}+{\bar\beta}+\alpha\beta+
\bar\alpha\bar\beta}$, in the type-II
construction, while one must set the statistical factor
$(-)^{\alpha+\beta+\alpha\beta}$=$+1$ (or $-1$) in the heterotic
constructions.
In order to derive the lower-mass levels we need the  behavior of
the bosonic and fermionic part of the partition function in the limit
where ${\rm Im}\tau$ is large (${\rm Im}\tau \rightarrow \infty$).
This behavior can be easily derived from (\ref{het}),
\begin{equation}
Z^{W}(\mu;\tau,{\bar\tau})\longrightarrow {\rm C}[{\rm
Im}\tau]^{-1}~e^{-\frac{{\rm Im}\tau}{2(k+2)}}.
\label{lomass}
\end{equation}
The above behavior is universal and does not depend on the choice of
$K^{(6)}$ internal $N=(2,2)$ space. Only the multiplicity factor $C$
(positive for bosons and negative for fermions) depends on the
different constructions and it is always proportional to the number
of the lower-mass level states with mass $\mu^2/2=1/[2(k+2)]=Q^2/4$.
If we
replace the $W^{(4)}_k$ with any one of the other $N=4$ ${\hat c}=4$
spaces, $C^{(4)}_k$, $\Delta^{(4)}_k(A,V)$, we get identical infrared
mass shift $\mu$.

As we will see in the next section,  the induced mass $\mu$ acts as a
well-defined  infrared regulator for all the on-shell correlation
functions and in particular for the two-point function correlator
$\langle F^{a}_{\mu\nu}F_{a}^{\mu\nu}\rangle$ (and  $\langle
R_{\mu\nu\rho\sigma}R^{\mu\nu\rho\sigma}\rangle$) on the torus, which
is associated
to the one-loop string corrections on the  gauge coupling constant.

\section{Non-zero ${\bf F^{a}_{\mu\nu}}$ and  $
R_{\mu\nu}^{\rho\sigma}$ Background in Superstrings}
\setcounter{equation}{0}

Our aim is to define the  deformation of the two-dimensional
superconformal theory  which corresponds to a non-zero field strength
$F^{a}_{\mu\nu}$ and $R_{\mu\nu\rho\sigma}$
background\footnote{Magnetic backgrounds in closed string theory have
been also discussed in \cite{bk,rt,t}.}
and find the integrated  one-loop
partition function $ Z^{W}(\mu,F,\R)$,  where $F$ is by the
magnitude
of the field strength,
$F^2 \equiv \langle F^{a}_{\mu\nu}F_{a}^{\mu\nu}\rangle$ and $\R$ is
that of the curvature,  $\langle
R_{\mu\nu\rho\sigma}R^{\mu\nu\rho\sigma}\rangle=\R^2$.

\begin{equation}
Z^{W} [\mu,F,\R]=\frac{1}{V(W)} \int_{\cal F}
\frac{ d\tau d{\bar\tau} }{ ({\rm Im}\tau)^2 }
Z^{W}[\mu,F,\R;\tau,{\bar\tau}]
\label{intpart}
\end{equation}
where $V(W)$ is the volume  of the $W^{(4)}_k$ space; modulo the
trivial infinity which corresponds to the one non-compact dimension,
the remaining three-dimensional compact space is that of the
three-dimensional sphere.
In our normalization:
$$
V(SU(2)_k)=\frac{1}{8\pi} (k+2)^{\frac{3}{2}}
$$
so that it matches in the flat limit with the conventional flat space
contribution.

 In flat space, a small non-zero  $F_{\mu\nu}^a$ background gives
rise
to an infinitesimal deformation  of the 2-d $\sigma$-model action
given by,
\begin{equation}
\Delta S^{2d}(F^{(4)})=\int dzd{\bar z}\;F_{\mu\nu}^a[x^{\mu}
\partial_z x^{\nu}+\psi^{\mu}\psi^{\nu}]{\bar J}_a
\label{fdef}
\end{equation}
Observe that for $F^a_{\mu\nu}$ constant (constant magnetic field),
the left moving operator $[x^{\mu} \partial_z
x^{\nu}+\psi^{\mu}\psi^{\nu}]$ is not a well-defined $(1,0)$ operator
on the world sheet. Even though  the right moving Kac-Moody current
${\bar J}_a$ is a well-defined $(0,1)$ operator, the total
deformation
is not integrable in flat space. Indeed, the 2-d $\sigma$-model
$\beta$-functions are not satisfied in the presence of a constant
magnetic field. This follows from the fact that there is a
$non$-$trivial$ $back$-$reaction$ on the gravitational background due
the non-zero
magnetic field.

The important property of $W^{(4)}_k$ space is that we can solve this
back-reaction ambiguity. First observe that the deformation that
corresponds to a constant magnetic field
$B_i^a=\epsilon_{oijk}F_a^{ik}$ is a well-defined
(1,1) integrable deformation, which breaks the $(2,2)$ superconformal
invariance but preserves the $(1,0)$ world-sheet supersymmetry:
\begin{equation}
\Delta S^{2d}(W^{(4)}_k)=\int dzd{\bar
z}\;B^a_i[I^i+\frac{1}{2}\epsilon^{ijk}\psi_{j}\psi_{k}]{\bar J}_a
\label{fdef1}
\end{equation}
where $I^i$ is anyone of the $SU(2)_{k}$ currents.
The deformed partition function is not zero due to the breaking of
$(2,2)$ supersymmetry.
In order to see that this is the correct replacement of the Lorentz
current in the flat case, we will write the SU(2) group element as
$g=\exp[i{\vec\sigma}\cdot{\vec x}/2]$ in which case
$I^{i}=kTr[\sigma^{i}
g^{-1}\p g]=ik(\p x^{i}+\epsilon^{ijk}x_j\p x_k+{\cal{O}}(|x|^3))$.
In the flat limit the first term corresponds to a constant gauge
field
and thus pure gauge so the only relevant term is the second one which
corresponds to constant magnetic field in flat space.
The $\cal{R}$ perturbation is
\be
\Delta S({\cal{R}})=\int dzd\bar
z\;{\cal{R}}\left[I^{3}+\psi^{1}\psi^{2}\right]
\bar I^{3}
\ee
In $\sigma$-model language, in the flat limit it gives the following
metric perturbation
\be
\delta (ds^2)=-{\cal{R}}\left[x^{1}dx^{2}-x^2dx^{1}\right]^2
\ee
with constant Riemann tensor and scalar curvature equal to
$6{\cal{R}}$.
There is also a  non-zero antisymmetric tensor with
$H_{123}=2\sqrt{{\cal{R}}}$
and dilaton $\delta
\Phi={\cal{R}}\left[(x^1)^2+(x^2)^2+4(x^3)^2\right]/4$.

Due to the rotation invariance in $S^3$ we can choose
$B_i^a=F\delta_i^3$ without loss of generality. The vector
$B^{a}_{i}$
indicates the
direction in the gauge group space of the right-moving affine
currents.
Looking at the $\sigma$-model representation of this perturbation,
we can observe that the $F_{\mu\nu}$ of this background gauge field
is a monopole-like gauge field on $S^3$ and its lift to the tangent
space is constant. Thus at the flat limit of the sphere it goes to
the constant $F_{\mu\nu}$ background of
flat space.

The moduli space of the $F$ deformation is then given by the
$SO(1,n)/SO(n)$ Lorentzian-lattice boosts with $n$ being  the rank
of the right-moving gauge group. We therefore conclude that the
desired partition function $Z^{W} (\mu,F,\R=0)$  is given in terms
of the moduli of the $\Gamma(1,n)$ lorentzian lattice. The constant
gravitational background
$R^{ij}_{kl}=\R\epsilon^{3ij}\epsilon_{3kl}$ can also be
included exactly
by an extra boost, in which case the lattice becomes
$\Gamma(1,n+1)$.

Let us denote by  $\Q$ the fermionic lattice momenta associated to
the left-moving $U(1)$ current $\partial H=\psi^1\psi^2$, by $I$
the charge lattice of the left-moving $U(1)$ current associated to
the $I_3$ current of $SU(2)_k$,
by $\bar \Q$ the charge lattice of a right U(1) which is part of
the Cartan algebra of the non-abelian right gauge group and
by $\bar I$ the charge lattice of the right-moving $U(1)$ current
associated
to the $\bar I_3$ current of $SU(2)_k$.
In terms of these charges the undeformed partition
function can be written as
\be
Tr[\exp[-2\pi \rm{Im}\tau (L_{0}+\bar L_{0})
+2\pi i\rm{Re}\tau (L_{0}-\bar L_{0})]]
\end{equation}
where
\be
{\rm L}_{0}={1\over 2}\Q^2+{I^2\over k}+\cdots\;,\;\bar {\rm
L}_{0}={1\over 2}\bar
\Q^2+{\bar I^2\over k}+\cdots
\ee
where the dots stand for operators that do not involve $I,\bar
I,\Q,\bar \Q$.

The (1,1) perturbation that turns on a constant gauge field strength
$F$ as well
as a constant curvature $\R$ background
produces an O(1,2) 2-parameter boost in $O(2,2)$, acting on the
charge lattice
above,
which transforms ${\rm L}_{0}$ and $\bar {\rm L}_{0}$ to
\be
{\rm L}_{0}'={\rm L}_{0}+{\cosh\psi-1\over 2}\left( {(\Q+I)^2\over
k+2}+\left(\cos\theta{\bar I\over \sqrt{k}}
+\sin\theta {\bar \Q\over \sqrt{2}}\right)^2\right)
+
\ee
$$+\sinh\psi{(\Q+I)\over \sqrt{k+2}}\left(\cos\theta{\bar I\over
\sqrt{k}}
+\sin\theta{ \bar \Q\over \sqrt{2}}\right)
$$
and
\be
{\rm L}'_{0}-\bar {\rm L}'_{0}={\rm L}_{0}-\bar {\rm L}_{0}
\ee

The parameters $\theta$  and $\psi$ are related to the constant
background fields
$F$ and $\R$ by\footnote{The $k$-dependence is such that there is
smooth flat
space limit.}
\be
F={\sinh\psi\sin\theta\over\sqrt{2(k+2)}}\;\;,\;\;
\R={\sinh\psi\cos\theta\over
\sqrt{k(k+2)}}
\ee
so that
\be
{\rm L}_{0}'-{\rm L}_{0}=(\Q+I)\left(\R\bar I
+F{ \bar \Q}\right)+
\ee
$$+
{\sqrt{1+(k+2)(2F^2+k\R^2)}-1\over 2}\left({(\Q+I)^2\over
k+2}+{\left(\R\bar I+
+F\bar \Q\right)^2\over (2F^2+k\R^2)}\right)
$$

The first term is the standard perturbation while the second term is
the back-reaction necessary for conformal and modular invariance.
Expanding the partition function in a power series in $F,\R$
\begin{equation}
Z^{W}(\mu,F,\R)=\sum_{n,m=0}^{\infty}F^{n}\R^{m}Z_{n,m}^{W}(\mu)
\end{equation}
we can extract the integrated correlators $\langle F^n R^m\rangle
=Z_{n,m}$.
For $\langle R\rangle$, $\langle F^2\rangle$, $\langle FR\rangle$ and
$\langle
R^2\rangle$ we
obtain:
\be
Z_{0,1}^{W}(\mu)=-4\pi{\rm Im}\tau \langle(\Q+I)\bar I\rangle
\label{newt}
\ee
\be
Z_{2,0}^{W}=8\pi^2{\rm Im}\tau^2\left[ \langle (\Q+I)^2\rangle
\langle (\bar \Q)^2\rangle-{2\langle(\Q+I)^2\rangle+(k+2)\langle \bar
\Q^2\rangle
\over 8\pi{\rm Im}\tau}\right]
\ee
\be
Z_{1,1}^{W}(\mu)= 16\pi^2{\rm Im}\tau^2 \langle \bar I\bar \Q\rangle
\left[\langle (\Q+I)^2\rangle-{k+2\over 8\pi{\rm Im}\tau}\right]
\ee
\be
Z_{0,2}^{W}(\mu)=8\pi^2{\rm Im}\tau^2\left[ \langle (\Q+I)^2\rangle
\langle \bar I^2\rangle-{k\langle(\Q+I)^2\rangle+(k+2)\langle \bar
I^2\rangle
\over 8\pi{\rm Im}\tau}\right]
\ee

The charges $\Q,\bar \Q$ in the above formulae act in the respective
$\vartheta\left[^{\alpha}_{\beta}\right](\tau,v)$-functions as
differentiation with respect to $v$.
In particular $\Q$ acts in the $\vartheta [^{\alpha}_{\beta}]$ of
eqs. (\ref{type2}), (\ref{het}),
$I,\bar I$ act in the level-$k$ $\vartheta$-function present in
$\Gamma(SU(2)_{k})$ (due to the parafermionic decomposition), and
$\bar \Q$ acts on one of  the right $\bar \vartheta$-functions.

It is straitforward to generalize the formulae above to the case
where
there are several gauge groups.
These are generated by a collection of antiholomorphic currents $\bar
J^{i}$
generating simple or $U(1)$ current algebras.
We normalize them so that $\langle\bar J^{i}(z)\bar
J^{j}(0)\rangle=k_{i}\delta^{ij}/2z^2$. This fixes the normalization
of the quadratic Casimirs in the simple factors.
Then,
\be
\delta L_{0}=\delta \bar L_{0}=(\Q+I)(\R \bar I+F_{i}\bar J^{i})+
\ee
$$+{-1+\sqrt{1+(k+2)(k_{i}F_{i}^2+k\R^2)}\over 2}\left[
{(\Q+I)^2\over k+2}+{(F_{i}\bar J^{i}+\R\bar I)^2\over
k_{i}F_{i}^2+k\R^2}
\right]
$$
Expanding again in the background fields up to second order we obtain
\bs
\label{formu}
\be
\langle F_{i}\rangle =-4\pi {\rm Im}\tau \langle (\Q+I)\rangle\langle
\bar J^{i}\rangle
\ee
\be
\langle \R\rangle =-4\pi {\rm Im}\tau \langle (\Q+I)\rangle\langle
\bar I\rangle
\ee
\be
\langle F_{i}^2\rangle =8\pi^2{\rm Im}\tau^2\left[ \langle
(\Q+I)^2\rangle
\langle (\bar
J^{i})^2\rangle-{k_{i}\langle(\Q+I)^2\rangle+(k+2)\langle (\bar
J^{i})^2\rangle
\over 8\pi{\rm Im}\tau}\right]
\ee
\be
\langle \R^2\rangle =8\pi^2{\rm Im}\tau^2\left[ \langle
(\Q+I)^2\rangle
\langle \bar I^2\rangle-{k\langle(\Q+I)^2\rangle+(k+2)\langle \bar
I^2\rangle
\over 8\pi{\rm Im}\tau}\right]
\ee
\be
\langle \R F_{i}\rangle = 16\pi^2{\rm Im}\tau^2 \langle \bar I\bar
J^{i}\rangle
\left[\langle (\Q+I)^2\rangle-{k+2\over 8\pi{\rm Im}\tau}\right]
\ee
\be
\langle F_{i}F_{j}\rangle = 16\pi^2{\rm Im}\tau^2 \langle \bar
J^{i}\bar J^{j}\rangle
\left[\langle (\Q+I)^2\rangle-{k+2\over 8\pi{\rm Im}\tau}\right]
\ee
\es
where we should rermember that $k+2=1/\mu^2$.

Renormalizations of higher terms can be easily computed.
We give here the expression for an $F_{i}^4$ term,

$$
\langle F_{i}^4\rangle={(4\pi{\rm Im}\tau)^4\over 24}\langle\left[
(\Q+I)^4  (\bar J^{i})^4-{3\over 4\pi{\rm Im}\tau}(\Q+I)^2(\bar
J^{i})^2
(k_{i}(\Q+I)^2+(k+2)(\bar J^{i})^2)+
\right.
$$
\be
\left.+{3\over 4(4\pi{\rm Im}\tau)^2}[k_{i}(\Q+I)^2+(k+2)(\bar
J^{i})^2]^2-
{3k_{i}(k+2)\over 2(4\pi{\rm Im}\tau)^3}[k_{i}(\Q+I)^2+(k+2)(\bar
J^{i})^2]
\right]\rangle
\ee

\section{One-loop Corrections to the Coupling Constants}
\setcounter{equation}{0}

The term linear in $R$ provides us with the one loop renormalization
of Newton's constant. It is obvious from (\ref{newt}) that this
renormalization is zero to one-loop since the only term that might
contribute from the left is $\Q$ and $\langle \bar I\rangle=0$ due to
global $SU(2)$
symmetry.
Strictly speaking, what we have computed is the renormalization of a
linear combination of Newton's constant and the axion-dilaton kinetic
term.
However we can disentangle the two by turning on a general $C_{ij}
(\Q_{i}+I_{i})\bar I_{j}$ background. For general constant $C_{ij}$
this
 satisfies the string equations to leading order, and this is
sufficient for computing the first order expectation value
(\ref{newt}) relevant for the renormalization of Newton's constant.
For this more general perturbation we still have
\be
C_{ij} \langle(\Q_{i}+I_{i})\bar I_{j} \rangle=0
\ee
again due to the global $SU(2)$ symmetry.
The above implies that in all 4-D heterotic string models with flat
spacetime, Newton's  constant and the kinetic axion-dilaton terms do
not renormalize
at one-loop. This is true also for models where supersymmetry is
spontaneously broken at tree level provided the one-loop cosmological
constant is finite
(no tachyons).

This argument generalizes in an obvious way to higher loops due to
the vanishing of $\langle \bar I^{3}\rangle$ on any genus Riemann
surface. This implies
that
Newton's constant and the axion, dilaton kinetic terms are not
renormalized in perturbation theory for
heterotic
backgrounds with $N\geq 1$ spacetime supersymmetry.\footnote{We were
informed by Minahan and Nemeschansky
that they reached a similar conclusion at one loop.}
An amusing fact is that Newton's constant does get a finite one-loop
renormalization in the respective type-II backgrounds (from N=(1,1)
sectors\footnote{If we have higher spacetime supersymmetry the
renormalization is zero due to extra zero modes.} ).
There we have
\be
Z_{0,1}^{\rm type-II}=-2\pi{\rm Im}\tau\langle (\Q+I)(\bar \Q+\bar
I)\rangle
\ee
Only the $\Q\bar \Q$ term saturates the 2d fermionic zero modes in
this case and we obtain
the following finite result
\be
\langle R\rangle_{\rm one-loop}\sim \int_{{\cal F}}{d^2\tau\over {\rm
Im}\tau^2}{\Gamma(SU(2))\over
V(SU(2))}={\pi\over 3}\left(1+2{\mu^2\over M_{str}^2}\right)
\ee
We have defined $M_{str}^2$ to be the mass of the lowest lying
oscillator
state in the string spectrum ($M_{str}^2=1/\alpha'$).

We now focus on  the one-loop correction to the
gauge
couplings\footnote{Calculations similar in spirit for ``topological"
quantites have been done in \cite{wl}.}, which is proportional to
$Z_{2,0}^{W}(\mu)$.
We can use the Riemann identity to transform the sum over  the
($\alpha,\beta)$ $\vartheta$-function characteristics (with non-zero
$v$)
that appear in (\ref{type2}), (\ref{het})
\be
{1\over 2}\sum_{a,b=0}^{1}(-)^{\alpha+\beta+\alpha\beta}
\vartheta[^{\alpha}_{\beta}](v|\tau)
\vartheta[^{\alpha+h_1}_{\beta+g_1}](0|\t)
\vartheta[^{\alpha+h_2}_{\beta+g_2}](0|\t)
\vartheta[^{\alpha-h_1-h_2}_{\beta-g_1-g_2}](0|\t)
=\ee
$$
=\vartheta[^1_1](v/2|\t)\vartheta[^{1-h_1}_{1-g_{1}}](v/2|\t)
\vartheta[^{1-h_2}_{1-g_{2}}](v/2|\t)
\vartheta[^{1+h_1+h_2}_{1+g_{1}+g_{2}}](v/2|\t)
$$
In this representation the charge operators are derivatives with
respect to $v$.

We will focus for simplicity to heterotic $Z_{2}\times Z_{2}$
orbifolds.
In this case all the characteristics in eq. (\ref{het}) take the
values
$0,1$.
The only non-zero contribution appears when one of the pairs
$(h_i,g_i)$ of
twists  is $(0,0)$ and the rest non-zero. There are three sectors
where
two out of the four fermion $\vartheta$-functions depend only on
$v/2$; they give non-zero contribution only when both derivatives
with respect to $v$  act on them. We have in total three $N=2$
sectors; the $N=4$ and the  $N=1$ sectors give zero contribution in
$Z_{2,0}(\mu)$ for the  $Z_{2}\times Z_{2}$ orbifold model.
For other orbifold models there might exist  non-zero contributions
coming
from the N=1 sectors.
Using  the fact that the contribution to the partition function of
the twisted bosons cancels (up to a constant that is proportional to
the number of fixed points)  that of the twisted fermions, and  also
the identity $\vartheta'(0)/2\pi=\eta^3$, we obtain the following
formula for $Z_{2,0}(\mu)$:

\be
Z^{A}_{2,0}(\mu)=- \sum_{i=1}^{3} \int_{\cal F} {d\t
d\tbb\over {\rm Im}\t}{\Gamma(SU(2))\over
V(SU(2))}{\Gamma^{i}_{2,2}(T_{i},U_{i})\over \bar \eta^{24}}
\left[\bar \Q_{A}^2-{1\over 4\pi{\rm Im}
\t}\right]2\bar\Omega(\tbb)\label{univ}
\ee
where $A$ indicates the appropriate gauge group ($E_{8}$, $E_{6}$ or
$U(1)$),
$\bar \Q_{A}$ is the associated charge operator, normalized so that
it
acts as
${i\over \pi}{\p\over \p\tbb}$ on the $\vartheta$-functions and
$\bar\Omega=\bar\Omega_8\bar\Omega_6$
with
\def\vt{\bar\vartheta}
\be
\bar\Omega_8(\tbb)={1\over
2}\sum_{a,b=0}^{1}\bar\vartheta^8[^{a}_{b}]\;\;,\;\;
\bar\Omega_6(\tbb)={1\over 4}\left[
\vt^8_{2}(\vt^4_3+\vt_{4}^4)-\vt_4^8(\vt_3^4+\vt_2^4)+
\vt_3^8(\vt_2^4-\vt_4^4)\right]
\ee
Thus the one-loop corrected gauge coupling constant can be written as
\be
{16\pi^2\over g_{A}^2(\mu)}={16\pi^2\over
g_{A}^2(M_{str})}+Z^{A}_{2,0}(\mu)
\label{cc}
\ee
Eq. (\ref{univ}) applies to any 4-d symmetric orbifold string model,
the only things that change are the moduli contribution from
$\Gamma^i(T,U)$
and the specific form of $\bar\Omega$.
This formula differs from that of \cite{moduli,dfkz,ant} since it
includes the
so-called universal contribution.
In particular the back-reaction of gravity is included exactly and
contributes to the universal terms.
Taking differences between different gauge groups we obtain the
regularized
form of the result of \cite{moduli,dfkz,ant}.
This result will be also presented below.
The only difference from their formula is the replacement of the flat
space
contribution by $\Gamma(SU(2))/V(SU(2))$.
Our result is $explicitly$ $modular$ $invariant$ and $finite$ both in
the IR and UV.

Before we proceed further some general remarks are in order.
First it is obvious from (\ref{formu}) that N=4 sectors (having
4 zero modes) do not contribute to the renormalization of coupling
constants.
Each operator $\Q$ soaks up one zero mode. Since we have at most
$\Q^2$
in our expressions N=4 sectors give a vanishing result.
N=2 sectors have 2 zero modes so only terms that contain $\Q^2$
contribute.
In such a case formulae (\ref{formu}) simplify to
\bs
\be
\langle F_{i}\rangle_{N=2}=\langle \R\rangle_{N=2}=0
\ee
\be
\langle F_{i}^2\rangle_{N=2}=8\pi^2{\rm
Im\tau}^2\langle\Q^2\rangle\left[
\langle (\bar J^{i})^2\rangle-{k_{i}\over 8\pi{\rm Im}\tau}\right]
\ee
\be
\langle \R^2\rangle_{N=2}=8\pi^2{\rm
Im\tau}^2\langle\Q^2\rangle\left[
\langle \bar I^2\rangle-{k\over 8\pi{\rm Im}\tau}\right]
\ee
\be
\langle \R F_{i}\rangle_{N=2}=16\pi^2 {\rm
Im}\tau^2\langle\Q^2\rangle
\langle\bar I\bar J^{i}\rangle
\ee
\be
\langle F_{i}F_{j}\rangle_{N=2}=16\pi^2 {\rm
Im}\tau^2\langle\Q^2\rangle
\langle \bar J^{i}\bar J^{j}\rangle
\ee
\es
Formula (4.8b) has been derived previously \cite{ant}.
These formulae are also valid for N=1 sectors since the terms linear in
$\Q$ that could contribute come with $I_{3}$ whose expectation value
is zero.

In order to clearly see  how  the $W^{(4)}_{k}$ acts as an IR
regulator, it is convenient  to perform the summation on the spin
index $l$ of the $SU(2)$ characters. This sum can be done
analytically and one obtains the following surprising (and eventually
useful)  identity
\be
\Gamma(SU(2)_k)=\sqrt{\rm{Im}\t}{(k+2)^{3/2}\over 8\pi}\left[{\p
Z(R)\over \p R}|_{R^2=k+2}-{1\over 2}\left( R\to R/2\right)\right]
\label{su2}
\ee
where $Z(R)$ is the $\Gamma (1,1)$ lattice contribution of the torus:
\be
Z(R)= \sum_{m,n}\exp\left[ {i\pi\t\over 2}\left({m \over
R}+nR\right)^2-{i\pi\tbb\over 2} \left({m \over R}-nR\right)^2\right]
\ee
Notice that the derivative with respect to $R$ in (\ref{su2})
subtracts the
$(m,n)=(0,0)$ contribution which is responsible for the IR
divergence.
In particular we have that the infrared cutoff $\mu=1/R$.

Using (\ref{su2}), $Z_{2}(\mu)$ becomes,
\be
Z^A_{2}(\mu)=  \sum_{i=1}^{3} 2 \int_{\cal{F}}{d\t d\tbb\over {\rm
Im}\t^2}
{\rm Im}\t^{1\over 2}\left[Z'(R)|_{k+2}-{1\over
2}\left(R\to R/2\right)\right]
\left[{\rm Im}\t
\Gamma^i_{(2,2)}(T_{i},U_{i})\Sigma^{A}\right]\label{32}
\ee
The function $\Sigma^A$ depends on the gauge group in question
and its constant part is proportional to the $\beta$-function
contribution of the N=2 sectors, $b_{i}=C(g_{A})-T(R^{i}_{A})$.
For example, in the $E_8$ case it is given by
\be
\Sigma^{E_8}=-2{\bar\Omega_6\over \bar\eta^{24}}\left[{i\p\over
\pi\p\tbb}-{2\over \pi\rm{Im}\t}\right]\bar\Omega_8
\label{sigma}
\ee
The differential operator in (\ref{sigma}) acts as a
covariant derivative on modular forms.

Eq. (\ref{32}) is the final form for the complete string one-loop
radiative
correction
to the appropriate gauge couplings.
This result is finite and manifestly invariant under the target space
duality group that acts on the $T_{i},U_{i}$ moduli.
We see in particular that the (regulated) integrand in our case is
related
to the partition function of a (3,3) lattice at special values of the
(3,3) moduli.
The derivative with respect to the $R$ modulus is responsible for the
regulation of the IR.
In order to see this  we will evaluate the part of the
radiative correction coming from the low-lying states (massless in
the limit
$\mu\to 0$), which in the
unregulated case is responsible for the IR divergence.
This is achieved by replacing the (2,2) lattice contribution in eq.
(\ref{32}) by 1 and leaving apart for the moment the universal
contribution (which is IR finite):
\be
Z^{\rm massless}_{2}(\mu)=\left[\sum_{i=1}^{3}b_i\right]
\int_{\cal{F}}{d\t
d\tbb\over
{\rm Im}\t} {\rm Im}\t^{1 \over 2}\left[Z'(R)|_{k+2}-{1\over
2}\left(R\to R/2\right)\right]
\ee
As expected, $Z^{\rm massless}_{2}(\mu)$ turns out to be finite and
for small
$\mu$ behaves like
\be
Z^{m=\mu}_{2}(\mu)=
(b_1+b_2+b_3)[\log(M_{str}^{2}/2\mu^2) +2c_0]+...\label{c}
\ee
where the dots stand for terms vanishing in the limit $\mu\to 0$.
The constant $c_{0}$ can be computed exactly with the result
\be
c_{0}={3\over 2}-{1\over 2}\log(\pi/2)+{1\over 2}\gamma_{E}
-{3\over 4}\log(3)=0.738857...
\ee
Observe that in this example the N=2 $\beta$-function coefficients
add to the full N=1 $\beta$-function coefficient,
$b_{1}+b_{2}+b_{3}=3C(g_{a})-T(R_{a})$.
This is always the case in any string ground state where the
Green-Schwarz
duality anomaly cancellation is unnecessary \cite{dfkz,ant}.
The constant coefficient $c_{0}$, together with that of massive
states
$F(T_{i},U_{i})$ as well as the universal contribution
define unambiguously the string scheme and can thus be compared with
the field theory result (regularized in the IR in the same way as
above) in any UV scheme, for instance the conventional $\bar {DR}$.
Although this coefficient is small, one has to compute the parts left
over
including the moduli dependence.
In particular the universal contribution can be important.
We calculate here the universal contribution due to would be massless
states, e.g. the constant part of $\bar\Omega/\bar\eta^{24}$.
This is equal to
\be
{60\over \pi}\int_{\cal{F}}{d^2\t\over
{\rm{Im}\t^2}}\sqrt{\rm{Im}\t}\left[Z'(R)|_{k+2}-{1\over
2}\left(R\to R/2\right)\right]=20+{\cal{O}}(\mu)
\ee
This contributes to the coefficient $c_{0}$ in (\ref{c}) equal to
$1/3$ for $E_{8}$ and $-5/21$ for $E_{6}$.
It implies that a full calculation is necessary, namely the
contributions from all massive states, in order to find the exact
string scheme.

We will now evaluate the integrals (\ref{32}) over the torus moduli
space
in order to obtain the full one-loop corrections to the coupling
constants.

As seen previously the one loop corrections involve integrals of the
form:
\be
I(R,U,T)={\partial \over \partial R}\int_{{\cal F}}{d^2\tau\over
{\rm
Im}\tau^2}[\sqrt{{\rm Im}\tau}Z(R)][{\rm Im}\tau \Gamma_{2,2}(U,T)]
\left(\bar F_{1}+{1\over {\rm Im}\tau}\bar F_{2}\right)
\label{inte}\ee

Here $\bar F_{1}$, $\bar F_{2}$ are antiholomorphic functions
with the following expansions
$$\bar F_{1}=\sum_{m=m_{0}}^{\infty}C_{m}e^{-2\pi im\bar\tau}$$
$$\bar F_{2}=\sum_{m=\hat m_{0}}^{\infty}\hat C_{m}e^{-2\pi im\bar
\tau}$$
where $m_{0}=0$ in the case of gauge coupling constant corrections
and $m_{0}=-1$ for the ${\cal R}^2$
coupling constant correction. $\hat m_{0}$ is $-1$ in both cases.
Moreover $\bar F_{1}+\bar F_{2}/{\rm Im}\tau$ is modular invariant.

For example, for the case of the $E_{8}$ correction we have
\be
Z_{2,0}^{E_{8}}(\mu)=\sum_{i=1}^{3}[I(1/\mu,U_{i},T_{i})-
I(1/2\mu,U_{i},T_{i})]
\ee
with
\be
\bar F_{1}=-{2i\over \pi}{\bar \Omega_{6}\over \bar
\eta^{24}}\partial_{\bar\tau}\bar \Omega_{8}\;\;,\;\;\bar F_{2}=
{4\over \pi}{\bar\Omega_{6}\bar\Omega_{8}\over \bar\eta^{24}}
\ee

The integral (\ref{inte}) can be evaluated using the method of orbits
of the
modular
group in order to unfold the integration domain.
There are three contributions, that of the zero orbit $I_{0}$, the
non-degenerate orbits $I_{1}$ and the degenerate orbits $I_{2}$
The perturbative IR divergence exists in $I_{2}$ although there are
extra divergences at special points in target moduli space (e.g.
$T=U$, $T=U=i$ and $T=U=\rho$).

We obtain
\ba\label{i1}
{I_{0}\over {\rm Im}T}&=&\int_{{\cal F}}{d^2\tau\over {\rm
Im}\tau^2}\left(\bar F_{1}+\bar F_{2}/{\rm Im}\tau\right)+{\cal
O}(\mu^2)=\nonumber\\
&=&\left[{2\over \pi}\bar G_{2}\bar F_{1}+{2\over \pi^2}\bar G_{2}^2
\bar
F_{2}\right]_{\bar q^0 {\rm term}}+{\cal O}(\mu^2)=\\
&=&-{2\pi\over 3}(C_{0}-24C_{-1})+{2\pi^2\over 9}(\hat C_{0}-48\hat
C_{-1})+{\cal O}(\mu^2)\nonumber
\ea
$$
I_{1}=-2\sum_{k=1}^{\infty}\sum_{l=[m_{0}/k]}^{\infty}C_{kl}
\left[\log[1-e^{ -2\pi i(k\bar T-lU)}]+c.c.\right]-
$$
\be
-2\sum_{k=1}^{\infty}\sum_{l=[\hat m_{0}/k]}^{\infty}\hat
C_{kl}\left[
\log[1-e^{-2\pi i(k\bar T-lU)}]+\right.
\ee
$$
+\left.{1\over 4\pi{\rm Im}(kT+lU)}F(e^{-2\pi i(k\bar
T-lU)})+c.c.\right]+{\cal O}(\mu^2)$$
with
\be
\bar G_{2}={\pi^2\over 3}\left[1-24\sum_{n=1}^{\infty}{n\bar
q^{n}\over 1-\bar q^{n}}\right],
\ee
$F(x)$ is related to the dilogarithm function
\be
F(x)=\int_{0}^{x}{du\over u}\log(1-u)
\ee
and $[x]$ stands for the integer part of $x$.

Finally
\be
I_{3}={{\rm Im}U{\rm Im}T\over \pi}\left(1-\mu{\partial\over \partial
\mu}\right)
{\sum_{m,n}}'{\sum_{j,p}}'\,{\mu^2e^{-2\pi i mn/\mu^2}\over {\rm
Im}T\,\mu^2\,|j+Up|^2+{\rm
Im}U\,m^2} \times\label{i3}
\ee
$$\times \left[ C_{mn}+{{\rm Im}U\over
\pi}{\mu^2\hat C_{mn} \over  {\rm Im}T\,\mu^2\,|j+Up|^2+{\rm
Im}U\,m^2
}\right]
$$

The above results imply that for differences of gauge
couplings\footnote{We exclude here $U(1)$'s that can get enhanced at
special points.}the
regulated result is proportional to

\be
\Delta_{AB}\sim -4{\rm Re}\log\eta(T)+2\left[\pi{\rm Im}T\mu^2
{\sum_{j,p}}'\,\sinh^{-2}\left(\pi\mu\sqrt{{\rm Im}T\over {\rm
Im}U}|j+Up|\right)-{1\over 2}\left(\mu\to 2\mu\right)\right]
+{\cal O}(\mu^2)
\ee
We can expose the IR divergent part as well as the duality invariance
of the result above by approximating for small $x$
\be
{1\over \sinh^{2}x}\approx {1\over x^{2}(1+x^2/6)^2}\approx {1\over
x^2}-
{1\over x^2+3}
\ee
Then, we obtain
\be
\Delta_{AB}\sim \log{M_{str}^2\over
\mu^2}-\log\left[|\eta(T)\eta(U)|^4T_{2}U_{2}\right]+
\log{3e^{2\gamma_{E}}\over \pi^2}+{\cal O}(1)
\ee
where the ${\cal O}(1)$ piece is moduli independent.
A more careful control of the subleading terms is needed in order to
compute
this constant part
The moduli dependent part agrees with the result of ref.
\cite{moduli}.

Although the general formulae (\ref{i1})-(\ref{i3}) are rather
explicit, it will be useful
to cast them in a form where the $T,U$ target space duality is
manifest.

Before ending this section we give also the regularized one-loop
correction
to the $\R^2$ coupling:
$$
Z^{\rm grav}(\mu)=-\sum_{i}\int_{\cal F}{d^2\tau\over {\rm Im}\tau^2}
{\rm Im}\tau \Gamma_{2,2}(T_{i},U_{i})\left[\bar I^2 -{1-2\mu^2\over
8\pi\mu^2{\rm Im}\tau}\right]{\Gamma_{SU(2)}(\mu)\over
V_{SU(2)}(\mu)}
{2\bar\Omega\over \bar \eta^{24}}
$$
\be
=\sum_{i}\int_{\cal F}{d^2\tau\over {\rm Im}\tau^2}
{\rm Im}\tau \Gamma_{2,2}(T_{i},U_{i}){2\bar\Omega\over \bar
\eta^{24}}X(\mu)
\ee
where
\be
X(\mu)=\sum_{m,n\in Z}(-1)^{m+n+mn}\left[\left(4\pi
i\partial_{\bar\tau}
\log\bar\eta-{\pi\over {\rm
Im}\tau}\right)\left(1-{2\pi|m-n\tau|^2\over
4\mu^2{\rm Im}\tau}\right)+\right.
\ee
$$\left. +{4\pi^2(m-n\tau)^2\over 48\mu^4{\rm
Im}\tau^2}\left(3-{2\pi|m-n\tau|^2\over
4\mu^2{\rm Im}\tau}\right)\right]\exp\left[-{\pi|m-n\tau|^2\over
4\mu^2{\rm Im}\tau}\right]
$$

In general, the coupling constant corrections depend on several
scales, namely
the expectation values of the moduli (here $T_{i}$ and $U_{i}$) and
the infrared scale $\mu$. In the next section we will show that the
one-loop corrections
to the couplings obey differential equations which relate changes of
the various scales above.

\section{IR Flow Equations for Couplings}
\setcounter{equation}{0}

Once we have obtained the one-loop corrections to the coupling
constants we can observe that they satisfy scaling type flows.
We will present here IR Flow Equations (IRFE) for differences of
gauge couplings.

The existence of IRFE is due to differential equations satisfied by
the
lattice sum of an arbitrary (d,d) lattice,
\be
Z_{d,d}={\rm Im}\tau^{d/2}\sum_{P_{L},P_{R}}e^{i\pi\tau
P^{2}_{L}/2-i\pi\bar\tau P_{R}^2/2}
\label{partition}
\ee
where
\be
P_{L,R}^2=\vec n G^{-1}\vec n+2\vec mBG^{-1}\vec n+\vec
m[G-BG^{-1}B]\vec m\pm 2\vec m \cdot \vec n
\label{momentum}
\ee
$\vec m,\vec n$ are integer d-dimesional vectors
and $G_{ij}$ ($ B_{ij}$) is a real symmetric (antisymmetric) matrix.
$Z_{d,d}$ is $O(d,d,Z)$ and modular invariant.
Moreover it satisfies the following second order differential
equation\footnote{The special case for $d=2$ of this equation was
noted and used in \cite{moduli,ant}.}:
\be
\left[ \left(G_{ij}{\partial\over \partial G_{ij}}+{1-d\over
2}\right)^2
+2G_{ik}
G_{jl}{\partial^2\over \partial B_{ij}\partial B_{kl}}
-{1\over 4}-4{\rm
Im}\tau^2{\partial^2\over \partial \tau\partial \bar
\tau}\right] Z_{d,d}=0\label{irfe}
\ee

The equation above involves also the modulus of the torus $\tau$.
Thus it can be used to convert the integrands for threshold
corrections to differences of coupling constants  into
total
derivatives on $\tau$-moduli space.
Using the equation above for the $\Gamma_{1,1}(\mu)$ and
$\Gamma_{2,2}(T,U)$
lattices we can evaluate the $\tau$ integral and we are left with a
differential equation for the couplings
with respect to the $T,U$ moduli and the IR scale $\mu$ only.
To derive such an equation we start from eq. (\ref{32})
to obtain
\be
\Delta_{AB}\equiv {16\pi^2\over g_{A}^{2}}-{16\pi^2\over g^{2}_{B}}=
(b_{A}-b_{B})\int_{\cal F}{d^{2}\tau\over {\rm
Im}\tau^{2}}A(R)B(T,\bar T,U,\bar U)\label{difere}
\ee
with
\be
A(R)=2\sqrt{{\rm Im}\tau}\left[Z'(R)|_{k+2}-{1\over
2}\left(R\to R/2\right)\right]
\ee
\be
B(T,\bar T,U,\bar U)={\rm Im}\tau \Gamma_{2,2}(T,\bar T,U,\bar U)
\ee
Eq. (\ref{difere}) does not apply to $U(1)$'s that can get enhanced
at special points of the moduli.
The general equation (\ref{irfe}) translates to the following
equations for $A$
and $B$:
\be
{1\over 16}\left[\left({\partial\over \partial R}R\right)^2
-1\right]A(R)={\rm Im}\tau^{2}\partial_{\tau}\partial_{\bar
\tau}A(R)\label{eeq1}
\ee
\be
{\rm Im}T^{2}\partial_{T}\partial_{\bar T}B(T,\bar T,U,\bar U)=
{\rm Im}\tau^{2}\partial_{\tau}\partial_{\bar \tau}B(T,\bar T,U,\bar
U)\label{eeq2}
\ee
and a similar one with $T\to U$.
Using (\ref{eeq1}), (\ref{eeq2}) we obtain
\be
\left[\left({\partial\over \partial R}R\right)^2 -1-16{\rm
Im}T^{2}\partial_{T}\partial_{\bar T}\right] \Delta_{AB}=
16(b_{A}-b_{B})\int_{\cal
F}d^{2}\tau\left[\partial_{\tau}\left(B\partial_{\bar\tau}A\right)
-\partial_{\bar\tau}\left(A\partial_{\tau}B\right)\right]
\label{boun}
\ee
The righthand side in (\ref{boun}) is a total divergence in moduli
space, getting contributions only from $\tau\to i\infty$. However the
contribution
there is zero due to the IR cutoff (unlike the unregulated case).
Thus, using $R=1/\mu$,  eq. (\ref{boun}) becomes
\be
\left[\left(\mu{\partial \over \partial \mu}\right)^2-2\mu{\partial
\over \partial \mu}-16{\rm Im}T^2{\partial^2\over \partial T\partial
\bar T}\right]\Delta_{AB}=0\label{irfe2}
\ee
and we have also a similar one with  $T\to U$.

We strongly believe that such equations also exist for single
coupling constants using appropriate differential equations for
$(d,d+n)$ lattices.

Notice first that the IR scale $\mu$ plays the role of the RG scale
in the effective
field theory (see eqs. (\ref{cc}) and (\ref{c})):
\be
{16\pi^2\over g_{A}^{2}(\mu)}={16\pi^2\over
g_{A}^2(M_{str})}+b_{A}\log
{M^2_{str}\over \mu^2}+ F_{A}(T_{i})+{\cal O}(\mu^2/M^2_{str})
\label{betaa}
\ee
where the moduli $T_{i}$ have been rescaled
by $M_{str}$ so they are dimensionless.
Second, the IRFE gives a scaling relation for the moduli dependent
corrections.
Such relations are very useful for determining the moduli dependence
of the threshold corrections.
We will illustrate below such a determination, applicable to the
$Z_{2}\times Z_{2}$ example described  above.

Using the expansion (\ref{betaa}) and applying the IRFE (\ref{irfe2})
we obtain
\be
{\rm Im}T^2{\partial^2\over \partial T\partial\bar
T}(F_{A}-F_{B})={1\over 4}(b_{A}-b_{B})\label{eq2}
\ee
and a similar one for $U$.
This non-homogeneous equation has been obtained in \cite{moduli,ant}.

Solving them we obtain
\be
F_{A}-F_{B}=(b_{B}-b_{A})\log[{\rm Im}T{\rm Im}U]+f(T,U)+g(T,\bar
U)+{\rm cc}
\ee
If at special points in moduli space, the extra massless states are
uncharged
with respect to the gauge groups appearing in (\ref{eq2}) then the
functions
$f$ and $g$ are non-singular inside moduli space.
In such a case duality invariance of the threshold corrections
implies
that
\be
F_{A}-F_{B}=(b_{B}-b_{A})\log[{\rm Im}T{\rm
Im}U|\eta(T)\eta(U)|^4]+{\rm constant}
\ee
This is the result obtained via direct calculation in \cite{moduli}.

It is thus obvious that the IRFE provides a powerful tool in
evaluating
general threshold corrections as manifestly duality invariant
functions of the moduli.

\section{Further Directions}
\setcounter{equation}{0}

We have presented an IR regularization for string theory (and field
theory) induced by the curvature of spacetime as well as by
non-trivial  dilaton and axion fields.
This regularization preserves a form of spacetime supersymmetry and
gives masses to all massless fields (including chiral fermions) that
are proportional to the curvature.
In particular, the theory is IR finite also at special values of the
moduli with extra massless states\footnote{See ref. \cite{lust} for a
recent attempt
to take into account such points in the unregulated approach.}.

In the regulated string theory we can compute exactly the one-loop
effective action for arbitrarily large, constant, non-abelian
gauge and gravitational fields.
Using this result, among other things, we can compute unambiguously
the
string-induced one-loop threshold corrections to the gauge couplings
as functions of the moduli.

Another set of important couplings that we have not explicitly
addressed in this paper are the Yukawa couplings.
Physical Yukawa couplings depend on the k\"ahler potential and the
superpotential.
The superpotential receives no perturbative contributions and thus
can be calculated at tree level.
The K\"haler potential however does get renormalized so in order to
compute
the one-loop corrected Yukawa couplings we have to compute the
one-loop renormalization of the K\"ahler metric.
When the ground state has (spontaneously-broken) spacetime
supersymmetry the wavefunction renormalization of the scalars
$\phi_{i}$ are the same as those for their auxiliary fields $F_{i}$.
Thus we need to turn on non-trivial $F_{i}$, calculate their
effective action on the torus and pick the quadratic part
proportional to
$F_{i}\bar F_{\bar j}$.
This can be easily done using the techniques we developed in this
paper
since it turns out that the vertex operators \cite{atsen} for some
relevant $F$ fields are bilinears of left and right U(1) chiral
currents.
The explicit computations will presented elsewhere.

There are several open problems that need to be addressed in this
context.

Although we have obtained an explicit formula for threshold
corrections
more work is needed so that it is cast in form where all the duality
symmetries
are manifest.

The structure of higher loop corrections should be investigated.
A priori there is a potential problem, due to the dilaton, at higher
loops.
One would expect that since there is a region of spacetime where the
string coupling become arbitrarily strong, higher order computations
would be problematic. We think that this is not a problem in our
models, because
in Liouville models with N=4 superconformal symmetry (which is the
case we consider) there should be no divergence due to the dilaton at
higher loops.
However, this point need further study.
One should eventually analyze the validity of non-renormalization
theorems at higher loops \cite{ant} since they are of prime
importance for phenomenology.

Once the full one-loop coupling corrections are known, and in the
absence
of higher order (perturbative) moduli-dependent corrections, it might
be possible to implement the S-duality conjecture \cite{s,sen} in
order to obtain non-perturbative results \cite{sw} concerning the
effective field theory of string theory.

The consequences of string threshold corrections for low energy
physics
should be studied in order to be able to make quantitative
predictions.

Finally, the response of string theory to the magnetic backgrounds
studied in this paper should be analysed since it may provide with
useful clues concerning
the behavior of strings in strong background fields and/or
singularities.

\vskip 1cm

\centerline{\bf Acknowledgements}

We would like to thank L. Alvarez-Gaum\'e, I. Antoniadis,  Z. Bern,
J. Cornwall, A. Dabholkar,
W. Lerche, J. Minahan, K. Narain, J. Polchinski, A. Strominger and
N. Warner for questions, discussions and suggestions.
One of us (C.K.) was  supported in part by EEC contracts
SC1$^*$-0394C and SC1$^*$-CT92-0789.


\begin{thebibliography}{9}


\bibitem{cand}
P.~Candelas, G.~Horowitz, A.~Strominger and E.~Witten, \np {\bf B258}
(1985) 46.

\bibitem{orbifold} L.J. Dixon, J. Harvey, C. Vafa and E. Witten, \np
{\bf B261} (1985) 678;\\ {\bf B274} (1986) 285;\\
K.S.~Narain, M.H.~Sarmadi and C.~Vafa, \np {\bf 288} (1987) 551.

\bibitem{nar}
K.S.~Narain, \np {\bf B169} (1986) 41;\\
K.S.~Narain, M.H.~Sarmadi and E.~Witten, \np {\bf B279} (1987) 369;

\bibitem{llsmap} W. Lerche, D. L\"ust and A.N. Schellekens, \pl {\bf
B181} (1986) 71;\\ \np {\bf B287} (1987) 477.

\bibitem{abk4d} I. Antoniadis, C. Bachas and C. Kounnas, \np {\bf
B289} (1987) 87;\\
H. Kawai, D.C. Lewellen and S.H.-H. Tye, \np {\bf B288} (1987) 1.

\bibitem{llsmap} W. Lerche, D. L\"ust and A.N. Schellekens, \pl {\bf
B181} (1986) 71; \np {\bf B287} (1987) 477.

\bibitem{gepner} D. Gepner, \pl {\bf B199} (1987) 370;
\np {\bf B296} (1988) 757.

\bibitem{effcl}
E.~Witten, \pl {\bf B155} (1985) 151;\\
S.~Ferrara, C.~Kounnas and M.~Porrati, \pl {\bf B181} (1986) 263;\\
S.~Ferrara, L.~Girardello, C.~Kounnas and M.~Porrati, \pl {\bf B193}
(1987) 368;\\
I.~Antoniadis, J.~Ellis, E.~Floratos, D.V.~Nanopoulos and T.~Tomaras,
\pl {\bf B191} (1987) 96;\\
S.~Ferrara, L.~Girardello, C.~Kounnas and M.~Porrati, \pl {\bf B194}
(1987) 358;\\
M.~Cvetic, J.~Louis and B.~Ovrut, \pl {\bf B206} (1988) 227;\\
M.~Cvetic, J.~Molera and B.~Ovrut, \pr {\bf D40} (1989) 1140;\\
L.~Dixon, V.~Kaplunovsky and J.~Louis, \np {\bf B329} (1990) 27.


\bibitem{moduli}
V.S.~Kaplunovsky, \np {\bf B307} (1988) 145;\\
L.J.~Dixon, V.S.~Kaplunovsky and J.~Louis, \np {\bf B355} (1991)
649.

\bibitem{dfkz} J.-P.~Derendinger, S.~Ferrara, C.~Kounnas and
F.~Zwirner, \np {\bf
B372} (1992) 145 and \pl {\bf B271} (1991) 307;\\
G.~Lopez Cardoso and B.A.~Ovrut, \np {\bf B369} (1992)351;\\
S. Ferrara, C. Kounnas, D. L\"ust and F. Zwirner, \np {\bf B365}
(1991) 431.


\bibitem{ant} I. Antoniadis, K. Narain and T. Taylor, \pl {\bf B276}
(1991) 37;\\
I. Antoniadis, E. Gava and K. Narain \pl {\bf B283} (1992) 209;\\
\np {\bf B383} (1992) 93;\\
I. Antoniadis, E. Gava, K.S. Narain and T. Taylor, \np {\bf B407}
(1993) 706;\\
ibid. {\bf B413}  (1994) 162;\\
Northestern preprint, NUB-3084, hep-th/9405024.

\bibitem{n4kounnas} C. Kounnas, \pl {\bf B321} (1994) 26;
Proceedings of the  International ``Lepton-Photon
Symposium and Europhysics Conference on High Energy
Physics", Geneva, 1991, Vol. 1, pp. 302-306;
Proceedings of the International Workshop on ``String Theory, Quantum
Gravity and Unification of Fundamental Interactions", Rome, 21-26
September 1992,
CERN preprint, CERN-TH.6790/93.

\bibitem{ki} E. Kiritsis, in the Proceeding of the International
Europhysics Conf. on HEP, Marseille, 1993, Eds. J. Carr and M.
Perrotet.

\bibitem{worm} I. Antoniadis, S. Ferrara and C. Kounnas \np {\bf
B421}
(1994) 343.

\bibitem{kktopol} E. Kiritsis and C. Kounnas, \pl {\bf B331} (1994)
51

\bibitem{min} J. Minahan, Nucl. Phys. {\bf B298} (1988) 36.


\bibitem{gcsbr}
H.-P.~Nilles, \pl {\bf 115} (1982) 193 and \np {\bf B217} (1983)
366;\\
S.~Ferrara, L.~Girardello and H.-P.~Nilles, \pl {\bf B125} (1983)
457;\\
J.-P.~Derendinger, L.E.~Ib\'a\~nez and H.P.~Nilles, \pl {\bf B155}
(1985) 65;\\
M.~Dine, R.~Rohm, N.~Seiberg and E.~Witten, \pl {\bf B156} (1985)
55;\\
C.~Kounnas and M.~Porrati, \pl {\bf B191} (1987) 91.

\bibitem{ssbr}
C.~Kounnas and M.~Porrati, \np {\bf B310} (1988) 355;\\
S.~Ferrara, C.~Kounnas, M.~Porrati and F.~Zwirner,
\np {\bf B318} (1989) 75;\\
I. Antoniadis, C. Bachas, D. Lewellen and T. Tomaras, Phys.Lett.
{\bf B207} (1988) 441;\\
C.~Kounnas and B.~Rostand, \np {\bf B341} (1990) 641;\\
I.~Antoniadis, \pl {\bf B246} (1990) 377;\\
I.~Antoniadis and C.~Kounnas, \pl {\bf B261} (1991) 369.

\bibitem{fkpz}
S.~Ferrara, C.~Kounnas and F.~Zwirner,
Nucl. Phys. {\bf B429} (1994) 589; ERRATUM, ibid. {\bf B433} (1995)
255;\\
C.~Kounnas, I.~Pavel and F.~Zwirner, \pl {\bf B335} (1994) 403.

\bibitem{cw}
C. Callan and F. Wilczek, Nucl. Phys. {\bf B340} (1990) 366.

\bibitem{trieste} E. Kiritsis and C. Kounnas,
CERN preprint, CERN-TH.7471/94, hep-th/9410212, to appear in the
proceeding of the Trieste Spring School and Workshop, 1994.

\bibitem{kikoulu} E. Kiritsis, C. Kounnas and D. L\"ust, Int. J. Mod.
Phys.
{\bf A9} (1994) 1361.

\bibitem{kac}V. G. Ka\v c, D. H. Peterson, Adv. in Math. {\bf 53}
(1984) 125.

\bibitem{nonc} K. Sfetsos, Phys.Lett. {\bf B271} (1991) 301;\\
I. Bakas and E. Kiritsis, Int. J. Mod. Phys. {\bf A7}
[Supp. A1] (1992) 55.

\bibitem{kprn4} C. Kounnas, M. Porrati and B. Rostand, \pl {\bf B258}
(1991) 61.

\bibitem{wormclas} C. Callan, J. Harvey and A. Strominger,
\np {\bf B359} (1991) 611;\\
C. Callan, Lectures at Sixth J.A. Swieca Summer School, Princeton
preprint PUPT-1278, (1991).

\bibitem{su2ch} D. Gepner and Z. Qiu, Nucl. Phys. {\bf  B285} (1987)
423.

\bibitem{bk} C. Bachas and E. Kiritsis, Phys. Lett. {\bf B325} (1994)
103.

\bibitem{rt} J. Russo and A. Tseytlin, CERN preprint,
CERN-TH.7494/94,
hep-th/9411099.

\bibitem{t} A. Tseytlin, Phys. Lett. {\bf B346} (1995) 55.

\bibitem{wl} W. Lerche, B. Nilsson, A. Schellekens and N. Warner,
Nucl. Phys. {\bf B299} (1988) 91;\\
W. Lerche, Nucl. Phys. {\bf B308} (1988) 102.

\bibitem{lust} G. Lopes Cardoso, D. L\"ust and T. Mohaupt, Humbolt
preprint,
HUB-IEP-94/50, hep-th/9412209.

\bibitem{atsen} J. Attick, L. Dixon and A. Sen, Nucl. Phys. {\bf
B292} (1987) 109.

\bibitem{s} A. Font, L. Ib\~anez, D. L\"ust and F. Quevedo,
Phys.Lett. {\bf B249} (1990) 35.

\bibitem{sen} A. Sen, Phys. Lett. {\bf B329} (1994) 217.

\bibitem{sw} N. Seiberg and E. Witten, Nucl. Phys. {\bf B426} (1994)
19; Erratum,  ibid.  {\bf B430} (1994) 485.

\end{thebibliography}
\end{document}